\documentclass[graybox]{svmult}

\usepackage{mathptmx}       
\usepackage{helvet}         
\usepackage{courier}        
\usepackage{type1cm}        
\usepackage{makeidx}         
\usepackage{graphicx}        
\usepackage{multicol}        
\usepackage[bottom]{footmisc}

\makeindex             

\begin{document}

\title*{MYStIX First Results: Spatial Structures of Massive Young Stellar Clusters}
\titlerunning{The Structure of YSCs}  
\author{Michael A. Kuhn, Adrian Baddeley, Eric D. Feigelson, Konstantin V. Getman, Patrick S. Broos, Leisa K. Townsley, Matthew S. Povich, Tim Naylor, Robert R. King, Heather A. Busk, Kevin L. Luhman, \& the MYStIX Collaboration}
\authorrunning{Michael A. Kuhn \& the MYStIX Collaboration} 
\institute{Michael A. Kuhn \at Department of Astronomy \& Astrophysics, Pennsylvania State University, 525 Davey Laboratory, University Park PA 16802, \email{mkuhn1@astro.psu.edu}
}
%
%
\maketitle

\abstract*{~~~~~ Observations of the spatial distributions of young stars in star-forming regions can be linked to the theory of clustered star formation using spatial statistical methods. The MYStIX project provides rich samples of young stars from the nearest high-mass star-forming regions. Maps of stellar surface density reveal diverse structure and subclustering. Young stellar clusters and subclusters are fit with isothermal spheres and ellipsoids using the Bayesian Information Criterion to estimate the number of subclusters. Clustering is also investigated using Cartwright and Whitworth's $Q$ statistic and the inhomogeneous two-point correlation function. Mass segregation is detected in several cases, in both centrally concentrated and fractally structured star clusters, but a few clusters are not mass segregated.}

\abstract{~~~~~ Observations of the spatial distributions of young stars in star-forming regions can be linked to the theory of clustered star formation using spatial statistical methods. The MYStIX project provides rich samples of young stars from the nearest high-mass star-forming regions. Maps of stellar surface density reveal diverse structure and subclustering. Young stellar clusters and subclusters are fit with isothermal spheres and ellipsoids using the Bayesian Information Criterion to estimate the number of subclusters. Clustering is also investigated using Cartwright and Whitworth's $Q$ statistic and the inhomogeneous two-point correlation function. Mass segregation is detected in several cases, in both centrally concentrated and fractally structured star clusters, but a few clusters are not mass segregated.}

\section{MYStIX: Spatial Distributions of Young Stars}



Spatial distributions of young stars in high-mass star-forming regions (HMSFR) may vary significantly in different regions (e.g.\ Cartwright \& Whitworth 2004) and provide information on the region's star-forming history and cluster dynamics (e.g.\ Parker et al.\ 2012). The Massive Young stellar cluster Study in the Infrared and X-ray (MYStIX) reveals a variety of star-cluster morphologies in its sample of nearby Galactic HMSFRs; a project overview is provided by Eric Feigelson in this volume. Sources in the X-ray ({\it Chandra}), near-IR (UKIDSS), and mid-IR ({\it Spitzer}) and published OB stars are probabilistically classified into disk-free, disk-bearing, and protostellar MYStIX Probable Cluster Members (MPCM).

Here we make a comparative investigation of structure in young stellar clusters using modern statistical methods to characterize the distribution of stars. Empirical trends from the comparison of regions may reveal underlying astrophysical phenomena. Targets include the Orion Nebula Cluster (ONC), W~40, NGC~2264, NGC~6334, NGC~6357, the Eagle Nebula, three clusters in the Greater Carina Cluster (Tr~14-15-16), and the Trifid Nebula, in approximate order of distance. These clusters have a variety of sizes, richnesses, and morphologies: the nearest, the ONC, is a single cluster with $\sim$3000 stars, while NGC~6357 is composed of three clusters each similar in richness to the ONC. For each region, the analysis includes $100-1000$ MPCM, and information about the total population may be inferred from the X-ray luminosity function (e.g.\ Getman et al. 2012).



\section{Parametric Cluster Modeling}

A variety of methods are used to map stellar distributions in clusters and identify subclusters (e.g.\ Gutermuth et al.\ 2009). Here we assume that the YSO population is made up of subclusters that may be described by parametric models. For young stellar clusters, surface densities may be modeled by isothermal spheres, or, more generally, ellipsoids. For example, Hillenbrand \& Hartmann (1998) successfully fit the ONC with an elliptical model $-$ this distribution function may be the result of dynamical relaxation; however, Galactic HMSFR are typically not old enough to have undergone two-body relaxation.

In statistical parlance, the collection of subcluster models is a finite mixture model; clustering properties are obtained through parameter estimation, and the number of clusters is determined by model selection. Models are fit by maximum-likelihood estimation (MLE) using the Bayesian Information Criterion, ${\rm BIC} = -2 \ln L + k \ln (n)$, where $L$ is the likelihood, $k$ is the number of \ parameters, and $n$ is the number of points. The minimum BIC is found through numerical optimization.

\begin{figure}
\centering
\includegraphics[angle=270.,width=4.3in]{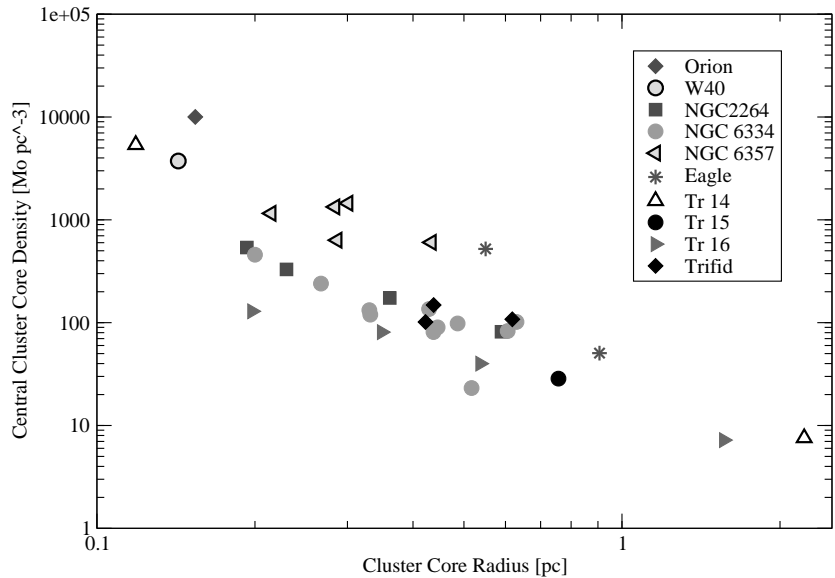}
\caption{Central densities and core radii for modeled subclusters in 10 MYStIX regions. Core radius is taken to be the average of the semi-major and semi-minor ellipse axes.
\label{fig1}}
\end{figure}

The ONC and W~40 are both fit by a single ellipsoid; the core size and shape of the ONC match Hillenbrand \& Hartmann's findings, and W~40 is roughly circular. Eleven components are found for NGC~6334, confirming the multiclusters of Feigelson et al. (2009). For NGC~6357, the furthest east cluster is fit by a single ellipsoid component, while the other two clusters are each fit by two components, confirming results of Wang et al. (2007). Figure 1 shows the negative correlation between subcluster core radius and central density for the modeled subclusters. This trend continues to hold true for subclusters in a single star-forming region. A similar trend was found by Pfalzner (2009), and may be related to subcluster age, with older clusters being less dense. The trend is slightly shallower than the $\rho \propto r^{-3}$ line. W~40 and the ONC are densest clusters, while Tr~15 is one of the least dense.

\section{Characterization of Subclustering}

Summary statistics, such as the two-point correlation function, have also been used to characterize spatial structure of young stellar clusters. However, these methods have difficulty distinguishing between first moment (gradients in surface density) and second moment (clustering) phenomena in point processes (Bartlett 1974). The $Q$ statistic (Cartwright \& Whitworth 2004) is cleverly designed to address this problem: it uses the mean edge length of the minimum spanning tree graph normalized by the average distance between points. The $Q$ parameter is calibrated through simulations so that fractal-like distributions of stars have $Q<0.8$, while centrally concentrated distributions of stars have $Q>0.8$. For MYStIX clusters the values of $Q$, in order of increasing central concentration, are: 0.61 for NGC~2264, 0.62 for NGC~6357, 0.71 for NGC~6334, 0.81 for Trifid, 0.82 for Tr~16, 0.83 for Tr~14, 0.89 for Orion, 0.9 for Tr~15, 0.91 for Eagle, and 0.92 for W~40. This agrees with the expectation that single clusters, like W~40, would have high central concentration, while a cluster with high sub-structure, like NGC~2264 would be more fractal.

This problem may also be addressed using the inhomogeneous two-point correlation function (Baddeley et al.\ 2000). In order to determine if the ellipsoid models are sufficient for describing the clustering of stars, the two-point correlation function is re-weighted by the subcluster mixture model, so that it will show clustering in the residuals that are not modeled. Values above the ``null hypothesis" line indicate additional clustering, and statistical significance may be evaluated using the 99\% confidence envelope. For the two cases above, the results indicate that the centrally concentrated W~40 cluster is adequately described by the ellipsoid model, while the model for NGC~2264 does not capture additional clustering at small separations.

\section{Mass Segregation}

Mass segregation of young stellar clusters may be related to both initial conditions of star formation and cluster dynamics. For the MYStIX sample, stellar masses are inferred from dereddened $J$ magnitudes, and spectroscopic catalogs of high-mass stars are taken from the literature.

A variety of methods have been used by astronomers to search for mass segregation in complex star-forming regions (e.g.\ K\"{u}pper et al.\ 2011). Here, we apply a second-moment statistic, $Emark$ (Schlather et al.\ 2004), designed to identify interaction between the position of a point and a value associated with it, here mass. The empirical $E(r)$ function gives the conditional expected value of mass when there is another star at projected distance $r$, and mass segregated clusters would have a decreasing slope. Statistical significance may be evaluated by generating a 99\% confidence envelope from simulations.

For the ONC, which is known to be mass segregated, this method shows a high significance detection. Mass segregation is also detected for W~40, confirming earlier results that assumed radial symmetry (Kuhn et al.\ 2010), and for NGC~2264, which has a fractal-like distribution of stars. Marginally significant mass segregation is found for NGC 6334, NGC 6357, and the Trifid Nebula. The Eagle Nebula shows no signs of mass segregation.

\vspace{5 mm}

The statistics of spatial point processes are used to compare young stellar clusters that have a diverse range of sizes, morphologies, and richnesses. This analysis reveals trends, such as the density-size relation, that may have implications for the astrophysics of star cluster formation and evolution. In a future study that includes the full MYStIX sample, we will seek patterns involving (sub)cluster populations, densities, sizes, mass segregation, fractality, and ages.

\vspace{-2 mm}

\begin{acknowledgement}
The MYStIX project is centered at Pennsylvania State University with the support of Chandra ACIS Team contract SV-74018, NSF AST-0809038, and NASA NNX09AC74G.   
\end{acknowledgement}

\biblstarthook{}

\end{document}